\documentclass[Physsubmission, Phys]{SciPost}

\hypersetup{
    colorlinks,
    linkcolor={red!50!black},
    citecolor={blue!50!black},
    urlcolor={blue!80!black}
}

\usepackage[bitstream-charter]{mathdesign}
\DeclareSymbolFont{usualmathcal}{OMS}{cmsy}{m}{n}
\DeclareSymbolFontAlphabet{\mathcal}{usualmathcal}

\newcommand{\numu}{\nu_\mu}
\newcommand{\numubar}{\bar{\nu}_\mu}
\newcommand{\nuebar}{\bar{\nu}_e}
\newcommand{\nue}{\nu_e}

\newcommand{\zmax}{z_{\rm max}}

\newcommand{\mdm}{m_\chi}
\newcommand{\Javg}{J_{\Delta\Omega}}
\newcommand{\sigmav}{\langle \sigma v \rangle}

\newcommand{\OmegaDM}{\Omega_{\rm DM,0}}
\newcommand{\Omegam}{\Omega_{m,0}}
\newcommand{\OmegaL}{\Omega_{\Lambda,0}}
\newcommand{\Mmin}{M_{\rm min}}

\newcommand{\Ekin}{E_{\rm kin}}
\newcommand{\Msun}{M_\odot}
\newcommand{\MeV}{{\rm \,MeV}}

\newcommand{\nuCraft}{{\tt nuCraft}}
\newcommand{\ROOT}{{\tt ROOT\,}}

\begin{document}
\begin{center}{\Large \textbf{
Disentangling Sub-GeV Dark Matter from the Diffuse Supernova
Neutrino Background using Hyper-Kamiokande\\
}}\end{center}

\begin{center}
Sandra Robles 
\end{center}

\begin{center}
Theoretical Particle Physics and Cosmology Group, Department of Physics, King’s College London, Strand, London, WC2R 2LS, UK
\\
sandra.robles@kcl.ac.uk
\end{center}

\begin{center}
\today
\end{center}


\definecolor{palegray}{gray}{0.95}
\begin{center}
\colorbox{palegray}{
  \begin{tabular}{rr}
  \begin{minipage}{0.1\textwidth}
    \includegraphics[width=30mm]{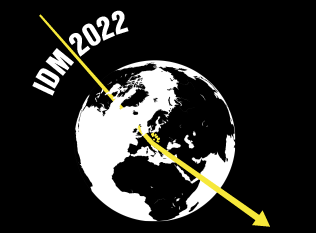}
  \end{minipage}
  &
  \begin{minipage}{0.85\textwidth}
    \begin{center}
    {\it 14th International Conference on Identification of Dark Matter}\\
    {\it Vienna, Austria, 18-22 July 2022} \\
    \doi{10.21468/SciPostPhysProc.?}\\
    \end{center}
  \end{minipage}
\end{tabular}
}
\end{center}

\section*{Abstract}
{\bf

The upcoming Hyper-Kamiokande (HyperK) experiment is expected to detect the Diffuse Supernova Neutrino Background (DSNB). This requires to ponder all possible  sources of background. 
Sub-GeV dark matter (DM) which annihilates into neutrinos is a potential background that has not been considered so far. We simulate DSNB and DM signals, as well as  backgrounds in the HyperK detector. We find that DM-induced neutrinos could indeed alter the extraction of the correct values of the parameters of interest for DSNB physics. Since the DSNB is an isotropic signal, and DM originates primarily from the Galactic centre, we show that this effect could be alleviated  with an on-off analysis.
}


\section{Introduction}
\label{sec:intro}
The Diffuse Supernova Neutrino Background (DSNB) is a steady state neutrino flux from all past core-collapse supernovae in the Universe. This quasi-thermal flux of ${\cal O}(10\MeV)$ is determined by  the effective temperature of the neutrinos emitted by the proto-neutron star and the supernova rate, which is related to the star formation rate (SFR). 
The DSNB has not been discovered yet, but it will be in the reach of the next generation neutrino telescopes, DUNE, JUNO and Hyper-Kamiokande (HyperK). 

Depending on the specific model the DSNB flux is expected to peak around $5\MeV$ and produce events below $50\MeV$. However, above $30\MeV$ atmospheric neutrinos overwhelm the DSNB signal and below $\sim16\MeV$ spallation products become an insurmountable background. In fact, 
the HyperK Design Report (DR)~\cite{Abe:2018uyc} assumes an energy window of $\sim16-30\MeV$ for DSNB analyses. On the other hand, it has been pointed out that 
HyperK  will have sensitivity to thermal dark matter (DM) annihilating into neutrinos in a similar energy range~\cite{Bell:2020rkw}. Therefore, we pose the question  whether or not  neutrinos from DM annihilation can contribute a significant background to DSNB searches. If so, is the DM signal sizeable enough to lead to incorrect inferences about the parameters of interest for DSNB physics, effective neutrino temperature and SFR? We show that this is indeed the case~\cite{Bell:2022ycf}. 

The paper is organised as follows. In section~\ref{sec:sigbkg} we outline the calculation of the DSNB and DM signals, as well as discuss sources of background. Our results are presented in section~\ref{sec:results} and concluding remarks are given in section~\ref{sec:conclusions}. 

\section{Signal and Background}
\label{sec:sigbkg}
We generate events for signal and background for 10 years of running time, using the fluxes calculated in this section and our original detector simulation~\cite{Bell:2020rkw}, which is based on \texttt{GENIE} neutrino Monte Carlo event generator v3.0.4a~\cite{Andreopoulos:2009rq,Andreopoulos:2015wxa}, the detector geometry was implemented using the \ROOT geometry package \cite{Brun:1997pa}. 

\subsection{Background}
Main backgrounds for DSNB searches include atmospheric neutrinos, invisible muons and spallation products. 
For atmospheric neutrinos, we use the FLUKA flux  below $100\MeV$ \cite{Battistoni:2005pd},  and the 4 dimensional HKKM11 flux \cite{Honda:2011nf} above $100\MeV$. Note that atmospheric neutrinos with $E_\nu\ge100\MeV$ also  contribute to the background in the energy window for DSNB searches, once events are binned in the final energy of the lepton $\Ekin$ \cite{Bell:2020rkw}. 
 In the absence of angular information for the FLUKA flux, we assume that  all their  energy bins  have the same angular distribution as that of the HKKM11 flux at $100\MeV$ \cite{Bell:2020rkw}.
 Both fluxes were oscillated using \nuCraft \cite{Wallraff:2014qka} and the  Preliminary Earth Reference Model~\cite{Dziewonski:1981xy}, with neutrino parameters from the Particle Data Group~\cite{Zyla:2020zbs}, assuming a normal mass ordering
as in ref.~\cite{Bell:2021esh}. 

Invisible muons are atmospheric $\numu$ and $\numubar$ that produce $\mu^-$ and $\mu^+$, respectively,  with kinetic energy below the threshold to produce Cherenkov photons. This muons decay to electrons or positrons that cannot be traced back to their parent neutrinos. We do not simulate this background, we take the expected event distribution from the HyperK DR~\cite{Abe:2018uyc}. 
Another important background at low energy are radioactive spallation products produced by cosmic ray muons that limit DSNB searches below $\sim16\MeV$~\cite{Abe:2018uyc}.

The detection channel for DSNB searches at water Cherenkov detectors  is the inverse beta decay of electron antineutrinos $\nuebar  + p \rightarrow e^+ + n$.  Since many of the background processes do not have a final state neutron, neutron tagging is a powerful method of background reduction. Neutron tagging  has been used in DSNB searches since the SuperK-IV analysis~\cite{Super-Kamiokande:2013ufi}, relying on the $2.2\MeV$ photon emitted after a direct neutron is captured by hydrogen. A more efficient method to tag neutrons relies on the addition of gadolinium salt to the water in the detector~\cite{Beacom:2003nk}. 
This technique allows to significantly reduce the invisible muon background. 
Similarly, charged current interactions of atmospheric $\nue$ can also be rejected. 

\subsection{Dark Matter Signal}
The main source of neutrinos from DM annihilation is the Galactic halo, more specifically the Galactic centre (GC). 
Following ref.~\cite{Bell:2020rkw}, 
the differential neutrino flux reads
\begin{equation}
\dfrac{{d\Phi_\nu}_{\Delta\Omega}}{dE_\nu} = \frac{\sigmav}{8\pi \mdm^2}  \Javg(a,z) \dfrac{dN_\nu}{dE_\nu} \, ,
\end{equation}
where $m_\chi$ is the DM mass, $\langle \sigma v \rangle$ is the thermally averaged annihilation cross section,  $J_{\Delta\Omega}(a,z)$ is the angle-averaged J-factor, and $dN_{\nu}/dE_{\nu}$ is the neutrino differential energy spectrum. Note that the J-factor is originally defined in galactic coordinates, we performed a coordinate transformation  to the horizontal $(a,z)$ coordinates defined at the detector location. This implies tracking the position of the GC in the sky, to account for this  we have averaged the J-factor over a 24 hours period as in Appendix B of ref.~\cite{Bell:2020rkw}.

A secondary source of neutrinos comes from extragalactic dark matter annihilation, integrated over redshift. This  diffuse isotropic neutrino flux  is given by~\cite{Beacom:2006tt}
\begin{equation}
\dfrac{d\Phi_\nu}{dE_\nu} = \frac{\sigmav}{2}\frac{c}{4\pi H_0}\frac{\OmegaDM^2\rho_{c,0}^2}{\mdm^2}\int_0^{z_{up}} dz \frac{(1 + G(z)) (1+z)^3}{\sqrt{\Omegam (1+z)^3+\OmegaL}}\frac{1}{3E_\nu}\delta\left[{z-\left(\frac{\mdm}{E_\nu}-1\right)}\right],
\end{equation} 
where $H_0$ is the Hubble constant, $\rho_{c,0}$ is the critical density of the Universe at $z=0$; $\OmegaDM$, $\Omegam$ and $\OmegaL$ are  the dark matter,  matter (baryonic and DM) and dark energy densities,  respectively (in units of $\rho_{c,0}$). 
 The halo boost factor $G(z)$  accounts for the enhancement to the annihilation rate due to  DM clustering in halos. It is computed by summing the contribution of all halos, 
i.e., a single halo contribution  
is weighted by the halo mass function (HMF). 
The HMF is calculated using the parametrization in ref.~\cite{Watson:2013}, assuming a minimum halo mass $\Mmin$. Since this value is not well constrained, we consider 
 two extreme cases as in ref.~\cite{Arguelles:2019ouk},  
 $\Mmin=10^{-3}\Msun$ (minimum) and $\Mmin=10^{-9}\Msun$ (maximum). 

\subsection{DSNB flux}

\begin{table}[t]
\centering
\begin{tabular}{|l|c|c|c|c|}
\hline
SFR analytic fits&  $ \dot{\rho}_0$ & $\alpha$  & $\beta$ & $\gamma$  \\
\hline
Upper & 0.0213 & 3.6 & -0.1 & -2.5 \\
Fiducial & 0.0178 & 3.4 & -0.3 & -3.5\\
Lower & 0.0142 & 3.2 & -0.5 & -4.5 \\
\hline
\end{tabular}
\caption{SFR density fits for the Salpeter IMF from ref.~\cite{Horiuchi:2008jz}.  }
\label{tab:SFRfits}
\end{table}  

The DSNB flux is obtained by redshifting the neutrino spectrum  $dN_{\nuebar}/dE_{\nuebar}$ from a single supernova (SN) according to the core collapse SN rate, which is calculated as the product of the star formation rate (SFR), $\dot{\rho}_\star(z)$, and the fraction of stars that end up their life cycles as neutron stars. Thus, the  DSNB flux reads
\begin{equation}
\dfrac{d\Phi_{\nuebar}}{dE_{\nuebar}}=\frac{c}{H_0}\int_0^{\zmax} dz \frac{1}{\sqrt{\Omegam(1+z)^3+\OmegaL}}  \dfrac{dN_{\nuebar}}{dE_{\nuebar}^{'}}(E_{\nuebar}^{'}) \dot{\rho}_\star(z) \frac{\int_8^{50} \xi(M) dM}{\int_{0.1}^{100} M \xi(M) dM} ,
\label{eq:DSNBflux}
\end{equation}
where  $\zmax=5$, $\xi(M)=M^{-2.35}$ is the Salpeter initial mass function (IMF)~\cite{Salpeter:1955it}, and 
\begin{equation}
\dfrac{dN_{\nuebar}}{dE_{\nuebar}^{'}}(E_{\nuebar}^{'})=\frac{120}{7\pi^4}\frac{E_\nu^{tot}}{6}\frac{E_{\nuebar}^{'2}}{T_{\nuebar}^4}\frac{1}{e^{E_{\nuebar}^{'}/T_{\nuebar}}+1}, 
\label{eq:nuebarspec}
\end{equation}
where $E_{\nuebar}^{'}=E_{\nuebar}(1+z)$, $T_{\nuebar}$ is the effective neutrino temperature outside the star after neutrino mixing, and $E_\nu^{tot}\simeq 3\times10^{53}\,$ erg~\cite{Horiuchi:2008jz} is the total neutrino energy, considering all flavours of neutrinos and antineutrinos. 
We assumed a continuous  broken power-law for the SFR~\cite{Yuksel:2008cu}, obtained by fitting Hubble and gamma-ray burst data in ref.~\cite{Horiuchi:2008jz}  
 \begin{equation}
  \dot{\rho}_\star(z) =  \dot{\rho}_0\left[ (1+z)^{\alpha\eta}  +  \frac{(1+z)^{\beta\eta}}{(1+z_1)^{(\beta-\alpha)\eta}} +  \frac{(1+z)^{\gamma\eta}}{(1+z_1)^{(\beta-\alpha)\eta} (1+z_2)^{(\gamma-\beta)\eta}} \right]^{1/\eta}, 
  \label{eq:SFRrho}
\end{equation}  
 where $\dot{\rho}_0$ is a normalisation constant in units $\Msun \, \rm yr^{-1} \, Mpc^{-3}$,  $\eta\simeq-10$~\cite{Yuksel:2008cu}, $z_1=1$, $z_2=4$, and  the logarithmic slopes of the low, intermediate, and high redshift regimes,  $\alpha$,  $\beta$ and $\gamma$ respectively are given in Table~\ref{tab:SFRfits}.

\section{Results}
\label{sec:results}

\begin{figure}[t]
\centering
\includegraphics[width=0.7\textwidth]{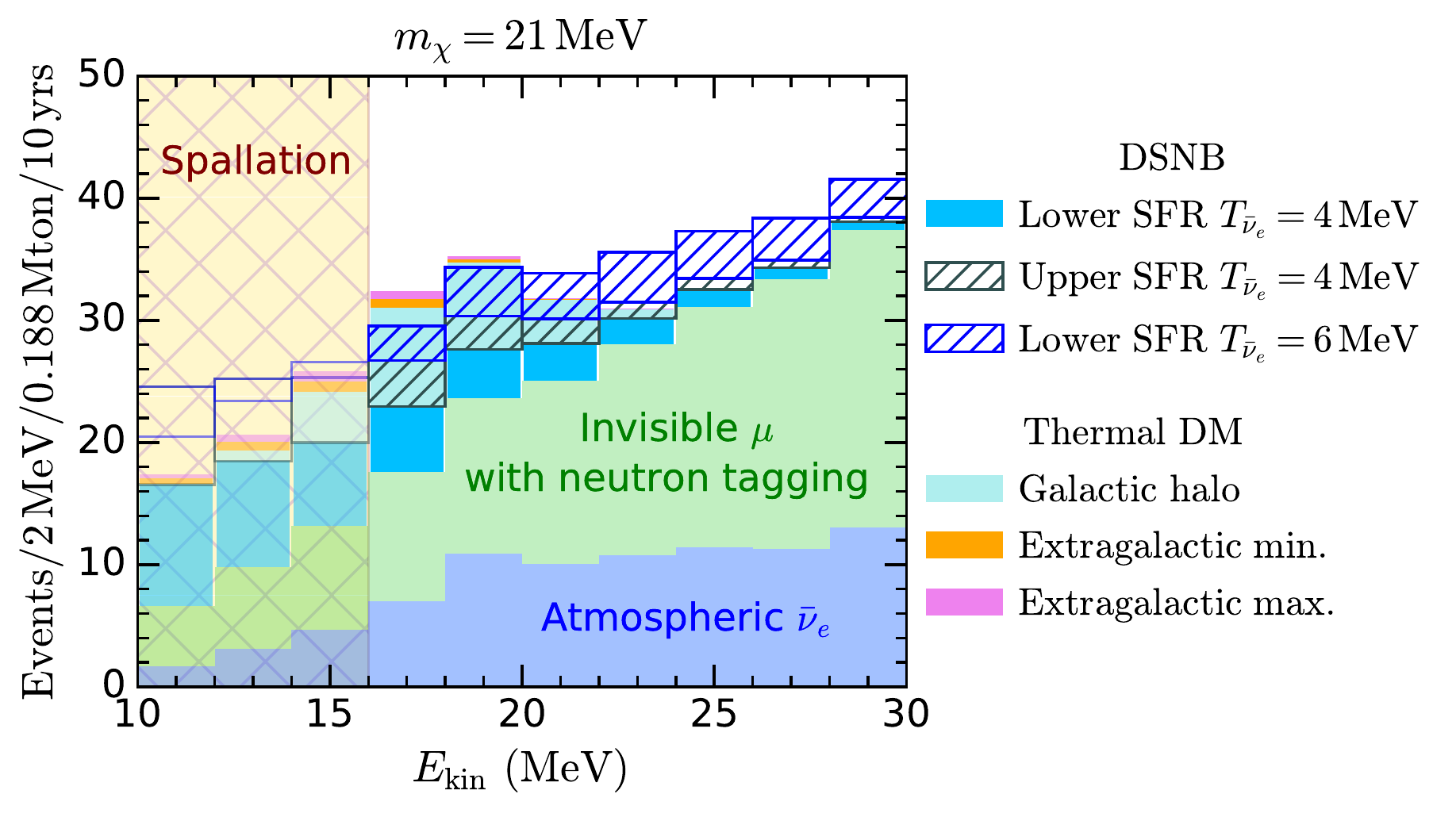}
\caption{Expected DSNB events for  the lower (light blue) and upper (hatched green) models with $T_{\nuebar}=4\MeV$, and the lower SFR fit with $T_{\nuebar}=6\MeV$ (hatched blue).
We also show thermal DM-induced  events, galactic (cyan) and extragalatic (orange and magenta), for  $m_\chi=21\MeV$;  as well as backgrounds from atmospheric neutrinos (purple), invisible muons assuming neutron tagging (green shaded),  and spallation products (hatched yellow). }
\label{fig:eventsntag}
\end{figure}

As in the Hyper DR, we consider an analysis window of $16-30\MeV$. 
To determine the impact a potential DM signal would have on  DSNB measurements, we consider nine DSNB models corresponding to the three SFR fits in Table~\ref{tab:SFRfits} (upper, fiducial and lower) with neutrino temperatures $T_{\nuebar}=4, \ 6$ and $8\MeV$, and compare different signals $S_i$ with and without neutrinos from DM annihilation using a test statistic from the profile log-likelihood ratio
\begin{equation}
TS= -2 \ln \frac{\mathcal{L}_p \left(\mathcal{D}_A (S_2)|S_1 \right)}{\mathcal{L}_p \left(\mathcal{D}_A (S_1)|S_2 \right)},
\label{eq:TS}
\end{equation}
where $\mathcal{D}_A$ is the Asimov dataset~\cite{Cowan:2010js}. We assume $TS\simeq\chi^2$ to estimate confidence intervals. 

As expected, DM of mass  $m_\chi\gtrsim30\MeV$ has no impact on the significance of the DSNB detection, since most of the events fall above the DSNB analysis window. 
 We find that the DM polluting effect is maximal for DM masses between 20 and $25\MeV$. For instance, suppose that $S_1$ in Eq.~\ref{eq:TS} is given by the lower SFR DSNB model with $T_{\nuebar}=4\MeV$ (light blue shaded), plus neutrinos from DM annihilation, galactic (cyan) and extragalactic components (orange and magenta for the minimum and maximum $\Mmin$ cases, respectively), for thermal DM with $\mdm=21\MeV$, 
 see Fig.~\ref{fig:eventsntag}. To form the log-likelihood ratio, we assume that $S_2$ is a different DSNB model without DM-induced neutrinos. We find that both the upper SFR model with $T_{\nuebar}=4\MeV$ (hatched green) and the lower model with $T_{\nuebar}=6\MeV$ (hatched blue) give equally good best fits, while the correct model would be ruled out at $\sim95\%$ CL. 
 From Fig.~\ref{fig:eventsntag}, note that
 neutrinos from DM of mass in the $20-25\MeV$ range do not contribute to the high energy bins, which have a lower statistical weight than those affected by DM pollution. This is because the backgrounds increase with increasing energy. 
 Thus, neutrinos from DM annihilation could indeed lead to quite misleading  interpretations of the DSNB dataset.  
 Note that we have assumed that the invisible muon background (green shaded) is significantly reduced by neutron tagging as estimated in the HyperK DR~\cite{Abe:2018uyc}, since we found that DSNB model discrimination with statistical significance would be impossible without n-tagging~\cite{Bell:2022ycf}. 
 
\begin{figure}[t]
\centering
\includegraphics[width=0.95\textwidth]{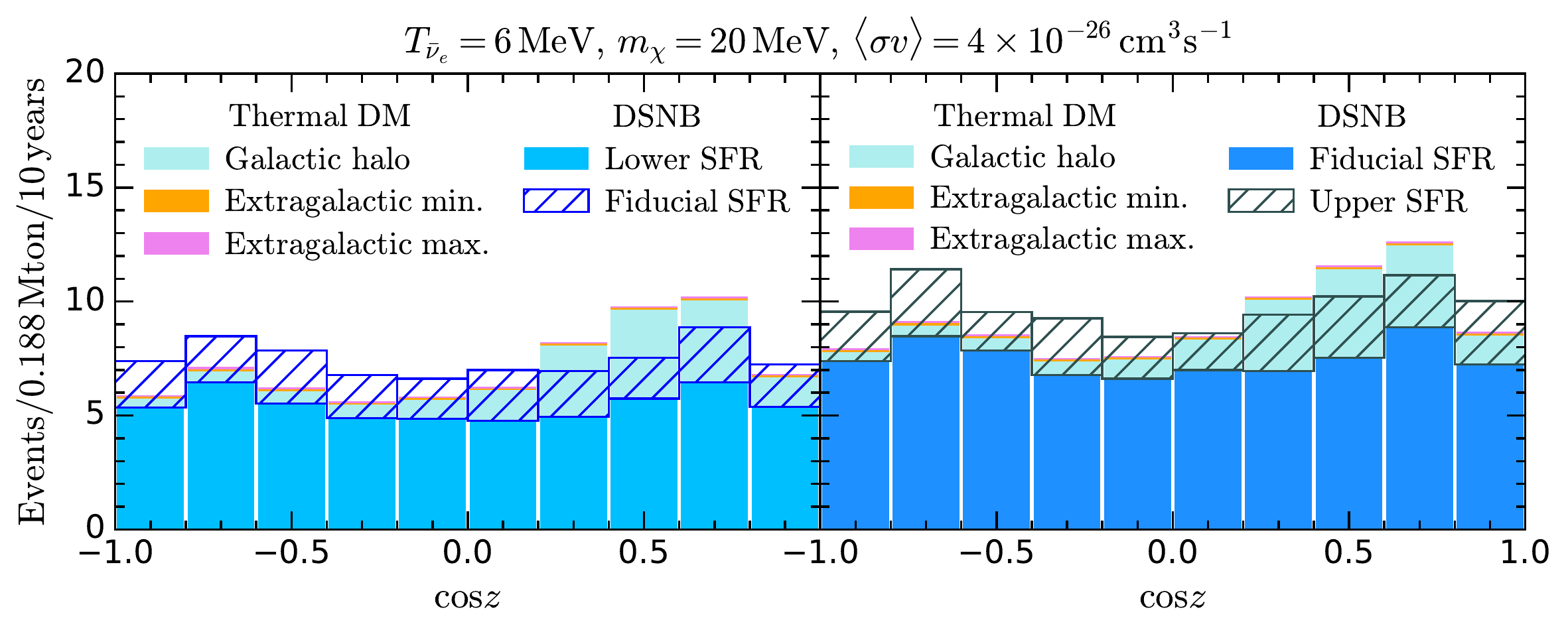}
\caption{DSNB and DM events binned in the zenith angle. The DM data corresponds to a thermal relic with mass $20\MeV$. The left panel compares the lower and fiducial SFR models with $T_{\nuebar}=6\MeV$ in the presence of  DM-induced galactic and extragalactic neutrino fluxes, while the right panel compares the fiducial and upper SFR DSNB models. }
\label{fig:eventscosz}
\end{figure}

A workaround to alleviate the effect of a  possible DM signal   would be the use of angular information in the DSNB analysis. The DSNB signal is isotropic up to statistical fluctuations, while the DM signal originates primarily from the GC. 
In Fig.~\ref{fig:eventscosz}, we show  DM and DSNB events  binned in the zenith angle ($\cos z$). As we can see, a simple on-off analysis that distinguishes between  positive and negative $\cos z$ would be enough to detect the presence of DM and recover the correct DSNB model in the off region for $T_{\nuebar}=6\MeV$.  Note, however, that there is a slight degeneracy between models when increasing the neutrino temperature and decreasing the SFR~\cite{Bell:2022ycf}. E.g. if the true DSNB parameters are the upper SFR and $T_{\nuebar}=6\MeV$, then the fiducial SFR model with $T_{\nuebar}=8\MeV$ is also a reasonable good fit.  
For $T_{\nuebar}=4\MeV$, most of the DSNB events fall below $16\MeV$ to allow the identification of the correct DSNB model. 

\section{Conclusion}
\label{sec:conclusions}
Next generation neutrino telescopes are expected to be sensitive to both the Diffuse Supernova Neutrino Background (DSNB) and light dark matter (DM) with thermal cross section. 
We have shown that sub-GeV DM which  annihilates predominantly to neutrinos can constitute a sizeable background for DSNB searches. Furthermore,  DM of mass in the $20-25\MeV$ range can lead to incorrect inferences about the star formation rate and the effective neutrino temperature, parameters of interests for DSNB physics. Only the use of angular information in the DSNB analysis  would enable to disentangle the DSNB and DM signals. 

\section*{Acknowledgements}
SR was supported by the UK STFC grant ST/T000759/1. 
SR  thanks the  Institute for Nuclear Theory at the University of Washington for its hospitality and the Department of Energy for partial support during the completion of this work.


\bibliography{references.bib}

\nolinenumbers

\end{document}